\documentclass[
letterpaper,
reprint,           
twocolumn,
superscriptaddress,
amsmath,           
amssymb,           
aps,               
prl,               
notitlepage,       
longbibliography,  
floatfix,          
nofootinbib,
]{revtex4-1}

\usepackage{amsmath}
\usepackage{lipsum}
\allowdisplaybreaks[4]
\usepackage{cancel}
\usepackage{extarrows}
\usepackage{tensor}     
\usepackage{float}
\usepackage[caption = false]{subfig} 
\usepackage[final]{graphicx}   
\usepackage[
colorlinks=true,        
citecolor=blue,         
linkcolor=blue,         
urlcolor=blue           
]{hyperref}             
\usepackage{bm}         
\usepackage{xcolor}     
\usepackage{lipsum}
\usepackage{color}      
\usepackage[utf8]{inputenc} 
\usepackage[section]{placeins} 
\usepackage{appendix}
\usepackage{units}
\newcommand{\nc}{\newcommand*}

\nc{\xbar}{\bar{x}}
\nc{\rhoeq}{\rho_{\mathrm{eq}}}
\nc{\zeq}{z_{\mathrm{eq}}}
\nc{\tla}{\tilde{\lambda}}
\nc{\bt}{\beta}
\nc{\dt}{\delta}
\nc{\Dt}{\Delta}
\nc{\vj}{\vec{j}}
\nc{\vl}{\vec{l}}
\nc{\hx}{\hat{x}}
\nc{\hy}{\hat{y}}
\nc{\bj}{\bm{j}}
\nc{\mJ}{\mathcal{J}}
\nc{\mP}{\mathcal{P}}
\nc{\Msun}{M_\odot}
\nc{\app}{\approx}
\nc{\av}[1]{\langle #1 \rangle}
\nc{\eq}[1]{Eq.~\eqref{#1}}
\nc{\al}{\alpha}
\nc{\Xstar}{X_{\ast}}
\nc{\fpbh}{f_{\mathrm{pbh}}}
\nc{\vth}{\vec{\theta}}
\nc{\vla}{\vec{\lambda}}
\nc{\vd}{\vec{d}}
\nc{\Mmin}{M_{\mathrm{min}}}
\nc{\rmd}{\mathrm{d}}
\nc{\mmin}{{m_{\mathrm{min}}}}
\nc{\mmax}{{m_{\mathrm{max}}}}
\nc{\mR}{\mathcal{R}}
\nc{\tmR}{\tilde{\mathcal{R}}}
\nc{\s}{\sigma}
\nc{\ogw}{\Omega_{\mathrm{GW}}}
\nc{\addref}{[\textcolor{red}{add ref}] }
\nc{\Om}{\Omega}
\nc{\gm}{\gamma}
\nc{\Gm}{\Gamma}
\nc{\gpcyr}{\mathrm{Gpc}^{-3}\,\mathrm{yr}^{-1}}
\nc{\Eq}[1]{Eq.~\eqref{#1}}
\nc{\Fig}[1]{Fig.~\ref{#1}}
\nc{\Table}[1]{Table~\ref{#1}}
\nc{\lvc}{LIGO/Virgo} 
\nc{\Sec}[1]{Sec.~\ref{#1}}
\nc{\eg}{\textit{e.g.~}}
\nc{\SNR}{\mathrm{SNR}}
\nc{\be}{\mathbf{\epsilon}}
\nc{\bn}{\mathbf{n}}
\nc{\bd}{\mathbf{d}}
\nc{\ba}{\mathbf{a}}
\nc{\eps}{\epsilon}
\nc{\bnu}{\mathbf{\nu}}
\nc{\mb}{\mathbf}
\nc{\bbt}{\mathbf{t}}
\nc{\bth}{\mathbf{\theta}}
\nc{\bep}{\mathbf{\epsilon}}
\nc{\uni}{\mathrm{U}}
\nc{\logu}{\operatorname{\mathrm{log-U}}}
\nc{\RN}{\mathrm{RN}}
\nc{\BN}{\mathrm{BN}}
\nc{\GN}{\mathrm{GN}}
\nc{\mcN}{\mathcal{N}}
\nc{\GWB}{\mathrm{GW}}
\nc{\yr}{\mathrm{yr}}
\nc{\Am}{\mathcal{A}}
\nc{\Dm}{\mathcal{D}}
\nc{\Hm}{\mathcal{H}}
\nc{\sovast}{Soviet Ast.}

\nc{\mrm}{\mathrm}
\nc{\BE}{B\scriptsize{AYES}\normalsize{E}\scriptsize{PHEM}\normalsize  }

\nc{\Ostgw}{\Omega_{\mathrm{GW}}^{\mathrm{ST}}}
\nc{\Ottgw}{\Omega_{\mathrm{GW}}^{\mathrm{TT}}}
\nc{\Ovlgw}{\Omega_{\mathrm{GW}}^{\mathrm{VL}}}
\nc{\Oslgw}{\Omega_{\mathrm{GW}}^{\mathrm{SL}}}
\nc{\cosxi}{\beta}

\def\({\left(}
\def\){\right)}
\def\[{\left[}
\def\]{\right]}

\def\e{\begin{equation}}
	\def\q{\end{equation}}
\def\m{\begin{eqnarray}}
	\def\n{\end{eqnarray}}
\nc{\red}[1]{\textcolor{red}{#1}}

\begin{document}

\title{Search for Stochastic Gravitational-Wave Background from Massive Gravity in the NANOGrav 12.5-Year Data Set}
\author{Yu-Mei Wu}
\email{wuyumei@itp.ac.cn} 
\affiliation{School of Fundamental Physics and Mathematical Sciences, Hangzhou Institute for Advanced Study, UCAS, Hangzhou 310024, China}
\affiliation{School of Physical Sciences, University of Chinese Academy of Sciences, No. 19A Yuquan Road, Beijing 100049, China}
\affiliation{CAS Key Laboratory of Theoretical Physics, Institute of Theoretical Physics, Chinese Academy of Sciences, Beijing 100190, China}

\author{Zu-Cheng Chen}
\email{Corresponding author: zucheng.chen@bnu.edu.cn}
\affiliation{Department of Astronomy, Beijing Normal University, Beijing 100875, China}
\affiliation{Advanced Institute of Natural Sciences, Beijing Normal University, Zhuhai 519087, China}

\author{Qing-Guo Huang}
\email{Corresponding author: huangqg@itp.ac.cn}
\affiliation{School of Fundamental Physics and Mathematical Sciences, Hangzhou Institute for Advanced Study, UCAS, Hangzhou 310024, China}
\affiliation{School of Physical Sciences, 
    University of Chinese Academy of Sciences, 
    No. 19A Yuquan Road, Beijing 100049, China}
\affiliation{CAS Key Laboratory of Theoretical Physics, 
    Institute of Theoretical Physics, Chinese Academy of Sciences,
    Beijing 100190, China}

\begin{abstract}
Gravitational waves offer a new window to probe the nature of gravity, including answering if the mediating particle, graviton, has a non-zero mass or not. Pulsar timing arrays measure stochastic gravitational wave background (SGWB) at $\sim1-100$~nanohertz. Recently, the North American Nanohertz Observatory for Gravitational Waves (NANOGrav) collaboration reported an uncorrelated common-spectrum process in their 12.5-year data set with no substantial evidence that the process comes from the SGWB predicted by general relativity. In this work, we explore the possibility of an SGWB from massive gravity in the data set and find that a massless graviton is preferred because of the relatively larger Bayes factor. Without statistically significant evidence for dispersion-related correlations predicted by massive gravity, we place upper limits on the amplitude of the SGWB for graviton mass smaller than $10^{-23}$~eV as $A_{\rm{MG}}<3.21\times 10^{-15}$ at $95\%$ confidence level. 
\end{abstract}
\maketitle

\textit{Introduction.} Endowing a non-zero mass to the graviton, which is a spin-2 massless particle mediating gravitational force suggested by general relativity (GR), has a long history that can be traced back to the work of Fierz and Pauli in the 1930s \cite{Fierz:1939ix}. The Fierz-Pauli theory is a linearized extension of GR, and it faces the van Dam–Veltman–Zakharov (vDVZ) discontinuity \cite{vanDam:1970vg, Zakharov:1970cc} in the massless limit. Moreover, most non-linear massive gravity theories are plagued with the Boulware–Deser ghost \cite{Boulware:1972yco}. The vDVZ discontinuity got resolved through the Vainshtein mechanism \cite{Vainshtein:1972sx} soon after its discovery, and several ghost-free realizations, such as the Dvali–Gabadadze–Porrati model \cite{Dvali:2000xg,Dvali:2000hr,Dvali:2000rv} and the de Rham-Gabadadze-Tolley model \cite{deRham:2010kj}, have been proposed during the past decades.

A massive graviton is expected to bring about several different effects compared with GR, and the graviton mass has been constrained by many gravitational experiments, such as the probe of Yukawa suppression of the Newtonian potential on the solar-system scale \cite{Bernus:2020szc} and on the large-scale of galactic clusters \cite{Goldhaber:1974wg}, the discrepancy between the observed and the expected decay rates of binary pulsar systems \cite{Finn:2001qi}, the null results in observing the superradiant instabilities in supermassive black holes \cite{Brito:2013wya}, and the weak lensing observation \cite{Choudhury:2002pu}. Note that some bounds are model-dependent and need to be taken with caution \cite{Will:1997bb}. One can refer to Ref.~\cite{deRham:2016nuf} for a summary of more bounds from current and future experiments.

With the thrilling direct detection of gravitational waves (GWs) \cite{LIGOScientific:2016aoc}, we have entered a new era where GWs can act as powerful tools in testing gravity, including the test on the dispersion relation of GWs during propagation and thus placing bounds on the graviton mass $m_g$. The first observed GW event GW150914 \cite{LIGOScientific:2016lio} has put an upper bound as $m_g \lesssim 1.2 \times 10^{-22}$~eV, and the bound has been continuing improved, for instance,  $m_g \lesssim 4.70 \times10^{-23}$~eV by GW Transient Catalog (GWTC)-1 \cite{LIGOScientific:2019fpa}, $m_g \lesssim 1.76\times 10^{-23}$~eV by GWTC-2 \cite{LIGOScientific:2020tif} and $m_g \lesssim 1.27\times 10^{-23}$~eV by GWTC-3 \cite{LIGOScientific:2021sio}. 

Besides the great success achieved by the ground-based GW detectors, the breakthrough in the detection of low-frequency GW is expected to be made by pulsar timing arrays (PTAs) within a few years.
By monitoring the times of arrival (ToAs) of radio pulses emitted by a set of millisecond pulsars over decades \cite{1978SvA....22...36S,Detweiler:1979wn,1990ApJ...361..300F}, PTAs are sensitive in the nanohertz band and optimal for searching for a stochastic gravitational-wave background (SGWB) via the correlation investigation \cite{Taylor:2015msb,Burke-Spolaor:2018bvk}. Specifically, the timing residuals induced by the SGWB predicted by GR will be encoded with the well-known Hellings $\&$ Downs correlations \cite{Hellings:1983fr} for widely spaced pulsars. However, in massive gravity theories, the correlated signature of timing residuals in PTAs induced by the SGWB will be different \cite{Lee:2010cg,Liang:2021bct}. It can therefore be used to study if the graviton has a non-zero mass or not.

Recently, one of the major PTA collaborations, North American Nanohertz Observatory for GWs (NANOGrav) \cite{McLaughlin:2013ira}, reported strong evidence for a stochastic common-spectrum process in their 12.5-year data set \cite{Arzoumanian:2020vkk}. This is a promising sign but still not enough to claim the detection of an SGWB because the evidence for the Hellings $\&$ Downs correlations is not significant in the data set. Similar results were also reported by the European PTA (EPTA) \cite{Kramer:2013kea}, the Parkes PTA (PPTA) \cite{Manchester:2012za}, and the International PTA (IPTA) \cite{Hobbs:2009yy} collaborations in their latest data sets \cite{Arzoumanian:2020vkk, Goncharov:2021oub,Antoniadis:2022pcn,Chen:2021rqp}.

As the origin of the common-spectrum process remains controversial \cite{Arzoumanian:2020vkk, Goncharov:2021oub}, several works have attempted to search for novel physics, such as non-tensorial polarizations \cite{Chen:2021wdo,Wu:2021kmd,Chen:2021ncc,NANOGrav:2021ini}, cosmological phase transitions \cite{NANOGrav:2021flc,Xue:2021gyq}, cosmic strings \cite{Chen:2022azo,Bian:2022tju} and ultralight dark matter \cite{PPTA:2021uzb,PPTA:2022eul} in the PTA data sets and have got some interesting results. In this letter, we will explore another possibility by adopting the NANOGrav 12.5-year data set to probe massive gravity. Finding no significant evidence for an SGWB from massive gravity, we place upper limits on the amplitude of the SGWB in the mass range of $[4\times 10^{-25}, 10^{-23}]$~eV.

\textit{Correlated timing residuals from an SGWB in massive gravity.} An SGWB causes delays in each pulsar's arrival time (or timing residuals) in a characteristic spatial correlated way. For the SGWB originated from a population of inspiraling supermassive black hole binaries, the corresponding induced cross power spectral density between any two pulsars, $a$ and $b$, can be modeled by a power-law form,
\e
S_{ab}(f) = \Gm_{ab}\frac{A^2}{12\pi^2}\left(\frac{f}{f_{\yr}}\right)^{-\gm}f_{\yr}^{-3},
\label{psd}
\q
where $A$ is the amplitude of the SGWB at the reference frequency $f_{\rm{yr}}=1/\rm{year}$, $\gamma$ is the spectral index which takes the value of $13/3$ \cite{Phinney:2001di}, and $\Gm_{ab}$ is the overlap reduction function (ORF) that describe the correlations between the pulsars as a function of pulsar pairs' angular separation. The ORF is crucial for detecting the SGWB predicted by GR or modified theories of gravity because it encodes rich information about the polarization and dispersion of gravity theories. We note that the uncorrelated common-spectrum process (UCP) reported by NANOGrav takes $\Gm_{\rm{UCP}}^{ab}=\delta_{ab}$.

Combining the de Broglie relations and the mass-energy equation, we describe the component GW signal of a massive SGWB with a four-wavevector $k^{\mu}=(k_0,\bf{k})$ that satisfies
\e
k_0=\sqrt{\frac{m_g^2 c^2}{\hbar^2}+\lvert\bf{k}\rvert^2},
\q
where $c$ is the speed of light, and $\hbar$ is the reduced Planck constant, the ORF takes \cite{Liang:2021bct}
\e
\begin{split}
\Gm_{\rm{MG}}^{ab} = & \frac{1}{16 \rm{\kappa}^5}
 \bigg [2\rm{\kappa}(3+(6-5\rm \kappa^2)\cosxi)\\
 &-6 \left(1+\cosxi +\rm{\kappa}^2(1-3\cosxi )\right)\ln \left(\frac{1+\rm{\kappa}}{1-\rm{\kappa}}\right)\\ 
&\left.-\frac{3 \left(1+2 \rm{\kappa}^2(1-2\cosxi)-\rm{\kappa}^4 (1-\cosxi^2 ) \right) \ln L}{ \sqrt{(1-\cosxi ) \left(2-\rm{\kappa}^2 (1+\cosxi) \right)}}\right],
\end{split}
\label{orf2}
\q
with $\cosxi \equiv \cos \xi$. Here
\e
\begin{split}
L=&\frac{1}{\left(\rm{\kappa}^2-1\right)^2}\bigg[1+2 \rm{\kappa}^2(1-2\cosxi)-\rm{\kappa}^4 (1-2\cosxi^2)\\
&-2 \rm{\kappa}(1-2 \rm{\kappa}^2 \cosxi )  \sqrt{(1-\cosxi ) \left(2-\rm{\kappa}^2 (1+\cosxi) \right)}\bigg],
\end{split}
\q
and $\rm{\kappa}$ is defined as
\e
{\rm{\kappa}} \equiv \frac{\lvert{\bf{k}}\rvert}{k_0}=\sqrt{1-\frac{f_{\rm{cut}}^2}{f^2}},
\label{kappa}
\q
with the mass-related cut-off frequency $f_{{\rm{cut}}} \equiv m_{g} c^2/(2\pi\hbar)$. We emphasize that only the two helicity-2 polarization modes are considered to obtain the above ORF. We do not consider the non-tensorial modes because, for massive gravity, the additional two helicity-1 polarization modes are unlikely to be produced in a natural physical process, and one helicity-0 mode is expected to get suppressed due to the Vainshtein screening mechanism \cite{deRham:2014zqa}. 

\begin{figure*}[htbp!]
	\centering
	\includegraphics[width=\textwidth]{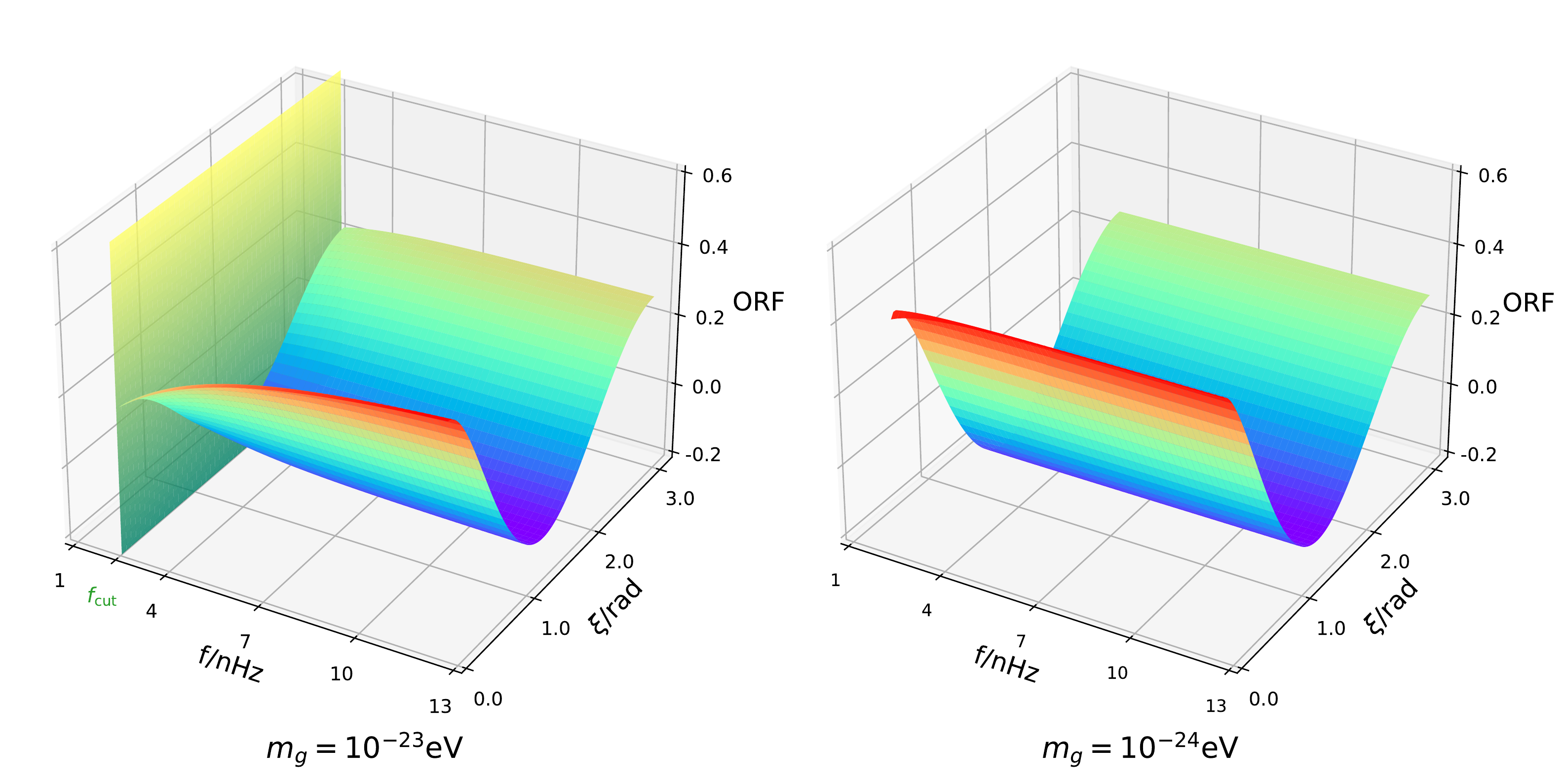}\caption{\label{ORF} ORF for the SGWB in massive gravity as a function of the GW frequency $f$ and the angular separation $\xi$. In particular, the frequency is chosen to range within the first 5 frequency bins of the NANOGrav 12.5-year data set, i.e., $f \in [1/T,5/T]=\unit[[2.5,12.6]]{\rm{\,nHz}}$ with $T$ the observational timespan. \textbf{Left panel:} the case of graviton mass $m_g=10^{-23}$~eV with the cut-off frequency $f_{\rm{cut}}=2.4 \rm{\,nHz}$ indicated by the green vertical plane on the f-axis. \textbf{Right panel:} the case of graviton mass $m_{g}=10^{-24}$~eV.}
\label{fig_ORF}
\end{figure*}

In the massless limit where ${\rm{\kappa}}=\lvert{\bf{k}}\rvert/k_0=1-\tilde{\epsilon}$ with $\tilde{\epsilon} \ll 1$, the ORF is approximated as
\e
\begin{aligned}
\Gm_{\rm{MG}}^{ab} \approx
&\frac{1}{8} \left(3+\cosxi +6 (1-\cosxi )\ln \frac{1-\cosxi}{2}\right) \\
&+\frac{\tilde{\epsilon}}{4}  \left(9+10\cosxi+(12-6\cosxi)\ln \frac{1-\cosxi}{2}\right).
\end{aligned}
\label{orf3}
\q
It reduces to the Hellings $\&$ Downs correlations when $\tilde{\epsilon}=0$\footnote{Note that in the massless limit, the ORF \eqref{orf3} is different from that given by Eq.~(40) in Ref.~\cite{Liang:2021bct}, which is not an explicit expansion for the parameter $\tilde{\epsilon}$.}.
From \Eq{orf2} and \Eq{kappa}, we see that $\Gm_{\rm{MG}}^{ab}$ depends on the graviton mass $m_{g}$, the GW frequency $f$, and the angular separation $\xi$ between two pulsars. We illustrate the ORF for two different graviton masses in \Fig{fig_ORF}.

\textit{Data Analysis.} The NANOGrav 12.5-year data set includes observations for 47 pulsars, of which 45 pulsars have an observational timespan over 3 years and have been used for the SGWB search \cite{Arzoumanian:2020vkk}. Here we follow Ref.~\cite{Arzoumanian:2020vkk} but exclude the PSR J0030$-$0451 because its credibility is in doubt in detection search due to the ill-modeled noise \cite{NANOGrav:2019ydy, NANOGrav:2021ini}.

In this work, we will search for the SGWB signal from massive gravity in the timing data. In practice, several effects also need to be accounted for to model the timing residuals properly. In particular, the massive-gravity SGWB effect should be analyzed along with the inaccuracies of the timing model, the measurement uncertainties of the timing, and the irregularities of the pulsar’s motion.
The observed pulse ToAs include several deterministic and stochastic effects. The expected arrival times are described by a timing model that characterizes the pulsar’s astrometric and timing properties, such as its position, proper motion, spin period, and additional orbital information if it is in a binary. Other stochastic contributions are from the uncorrelated process (white noise) and the correlated process (red noise). Following Ref.~\cite{Arzoumanian:2020vkk}, the timing residuals $\dt \bbt$ after fitting for the timing model can be decomposed as
\e
\dt \bbt=\dt \bbt_{\rm{TM}}+ \dt \bbt_{\rm{WN}}+\dt \bbt_{\rm{RN}}+\dt \bbt_{\rm{SGWB}},
\q
where $\dt \bbt_{\rm{TM}}$ accounts for the inaccuracy of the timing model, $\dt \bbt_{\rm{WN}}$ is the white noise term that accounts for measurement uncertainties, $\dt \bbt_{\rm{RN}}$ is the red noise from rotational irregularities of the pulsar, and $\dt \bbt_{\rm{SGWB}}$ is the SGWB signal we are searching for.
The inaccuracy of the timing model is given by 
\e
\dt \bbt_\mathrm{TM} = M \bep,
\q
where $M$ is the design matrix, and $\bep$ is an offset vector of timing model parameters. 
As the measurement errors might be underestimated, the white noise is modeled by a diagonal  covariance matrix $C^{\rm{WN}}$ with the modified uncertainties components,
\e
\sigma_{I,j}^2 =({\rm{EFAC}}\,\, \sigma _{I,j}^{\rm{ToA}})^2+\rm{EQUAD}^2,
\q
where $\sigma^{\rm{ToA}}_{I,j}$ is the formal $j$-th ToA uncertainties for the given pulsar $I$, EFAC is the scale factor that accounts for the possible miscalibration of radiometer noise in each observing ``system" that contains telescope, recording system and receiver \citep{Lentati:2015qwp}, and EQUAD is an additional term independent of uncertainties that is used to describe other source of time-independent noise, such as jitter noise that is varied across different pulsars \cite{Lentati:2015qwp, NANOGrav:2015qfw}. In addition, another parameter ECORR is also used to describe the ToA errors that are correlated within the same observing epoch but uncorrelated between different observing epochs \cite{NANOGrav:2015aud}. The red noise is modeled with a power-law spectrum with the amplitude $A_{\rm{RN}}$ and the index $\gm_{\rm{RN}}$,
\e
S(f)=\frac{A_{\rm{RN}}^2}{12\pi^2}\left(\frac{f}{f_{\yr}}\right)^{-\gm_{\rm{RN}}}f_{\yr}^{-3},
\q
and its covariance matrix is 
\e
C^{\rm{SN}}_{i,j}=\int d f \,S(f) \cos(2\pi f(t_i-t_j)),
\label{CM}
\q
where $t_i$ and $t_j$ are the $i$-th and $j$-th ToAs. The ``Fourier-sum" method is adopted to approximate the integral where 30 discrete frequency modes are chosen, i.e., $f={1/T, 2/T,...30/T}$ with $T$ the observational timespan. For the SGWB signal, the covariance matrix $C^{\rm{SGWB}}$ resembles that of the red noise \eqref{CM}, but with the power spectral density taking the form of \Eq{psd}. Besides, we take 5 frequency modes for the calculation of $C^{\rm{SGWB}}$ to reduce the potential coupling between the high-frequency components of the common process and the white noise \cite{Arzoumanian:2020vkk}.

\begin{table*}[!htbp]
    \centering
    \caption{Parameters and their prior distributions used in the analyses.}
    \label{prior}
    \begin{tabular}{c c c c}
        \hline
        \textbf{Parameter} & \textbf{Description} & \textbf{Prior} & \textbf{Comments} \\
        \hline
        \multicolumn{4}{c}{White Noise}\,\\	        
        $E_{k}$ & EFAC per backend/receiver system & $\uni[0, 10]$ & single pulsar analysis only \\
        $Q_{k}$[s] & EQUAD per backend/receiver system & $\logu[-8.5, -5]$ & single pulsar analysis only \\
        $J_{k}$[s] & ECORR per backend/receiver system & $\logu[-8.5, -5]$ & single pulsar analysis only \\
        \hline
        \multicolumn{4}{c}{Red Noise} \\
        $A_{\rm{RN}}$ & red-noise power-law amplitude &$\logu[-20, -8]$ & one parameter per pulsar\, \\
        $\gamma_{\rm{RN}}$ & red-noise power-law index  &$\uni[0,10]$ & one parameter per pulsar\, \\
        \hline
        \multicolumn{4}{c}{Common-spectrum Process}\,\\
        $A_{\mrm{UCP}}$ & UCP power-law amplitude &$\logu[-18, -14]$ & one parameter per PTA\, \\
        $A_{\mrm{MG}}$ & amplitude of SGWB from massive gravity  &$\logu[-18, -14]$ & one parameter per PTA\, \\
        $m_g [\rm{eV}]$ & graviton mass &\,\, delta function in $[10^{-24.4}, 10^{-23}]$ \,\,& \,\,$m_g \in \{10^{-24.4},10^{-24.3},...,10^{-23}\}$\, \\
        \hline
    \end{tabular}
\end{table*}

Assuming the stochastic processes are Gaussian and stationary \cite{Ellis:2013nrb}, the likelihood is evaluated by a multivariate Gaussian function, 
\e
L(\dt \bbt|\Theta, \{\eps_{n}\})=\frac{1}{\sqrt{(2\pi) {\rm{det}}(C)}}\exp\(-\frac{1}{2}\mb{r}^{\rm{T}}C^{-1}\mb{r}\)
\q
where $\mb{r}=[\dt \bbt_1 -M_1\bep_1, \dt \bbt_2 -M_2\bep_2, ..., \dt \bbt_{N} -M_{N}\bep_{N} ]^{\rm{T}}$ is a collection of $\dt \bbt -M\bep$ that accounts for the contribution from all of the stochastic noise and signals for all the $N$ pulsars, $C=\langle\mb{r}\mb{r}^{\rm{T}}\rangle$ is the total covariance matrix. In data analyses, we use the \texttt{TEMPO2} timing software \cite{Hobbs:2006cd,Edwards:2006zg} to determine the timing model design matrix $M$ and use the \texttt{Enterprise} package \cite{enterprise} to calculate the likelihood $L(\dt \bbt|\Theta)$ by marginalizing over the timing model uncertainty parameters $\bep$.

We infer the model parameter $\Theta$ by employing Bayes’ theorem
\e
P(\Theta|\dt \bbt)=\frac{L(\dt \bbt|\Theta)\pi(\Theta)}{Z},
\q
where $\pi(\Theta)$  is the prior probability distribution and $Z$ is the evidence given by the integral of  the likelihood times the prior over the prior volume,
\e
Z=\int L(\dt \bbt|\Theta)\pi({\Theta}) d{\Theta}.
\q
In analyses, we use the \texttt{PTMCMCSampler} package \cite{justin_ellis_2017_1037579} to conduct the Markov-chain Monte-Carlo sampling needed for parameter estimation. The parameters and their prior distributions needed for the analyses are given in \Table{prior}. 
Firstly, we perform the noise analyses by only including the white and red noise for every single pulsar. Then we collect all the 44 pulsars as a whole PTA, fix the white noise parameters to their maximum-likelihood values estimated from the single pulsar noise analyses, and allow red noise parameters to vary simultaneously with the SGWB signal parameters. In signal search among all the pulsars, fixing white noise parameters has negligible impact on the result \citep{Lentati:2015qwp}, but can efficiently reduce the computational cost.

For two possible candidates, $H_1$ and $H_0$, we employ the Bayes factor,
\e
\rm{BF}=\frac{Z_{1}}{Z_{0}},
\q
to measure which model fits the data better.
Usually,  $\rm{BF}>3$ can be interpreted as positive preference for $H_1$ over $H_0$, but only when $\rm{BF}>30$ can one declare a strong support for $H_1$ \cite{BF}. 
In practice, we use the product-space method \cite{10.2307/2346151,10.2307/1391010,Hee:2015eba,Taylor:2020zpk} to estimate the BFs, as was done in \cite{Arzoumanian:2020vkk}.

\textit{Results.} In data analyses, the covariance matrix is calculated at linearly spaced frequency modes $i/T$ ($i\in 1,2,3, \cdots$), which results in an upper detection limit on the graviton mass by the PTA because the cut-off frequency in \Eq{kappa} should not be larger than the inverse of the observational timespan, $1/T$. 
For the 12.5-year data set, the estimated upper limit is $m_g \lesssim 1.05\times 10^{-23}$~eV.
Note also that if $m_g \lesssim 4\times10^{-25}$~eV, we have $\rm{\kappa}>0.999$ in the whole frequency band. Therefore we take $4\times10^{-25}$~eV as a sufficient lower mass cut-off to approximate the massless limit, and search for the SGWB with the graviton mass in the range of $m_{g}\in [4\times 10^{-25},10^{-23}]$~eV.

\begin{figure}[htbp!]
	\centering
	\includegraphics[width=0.5\textwidth]{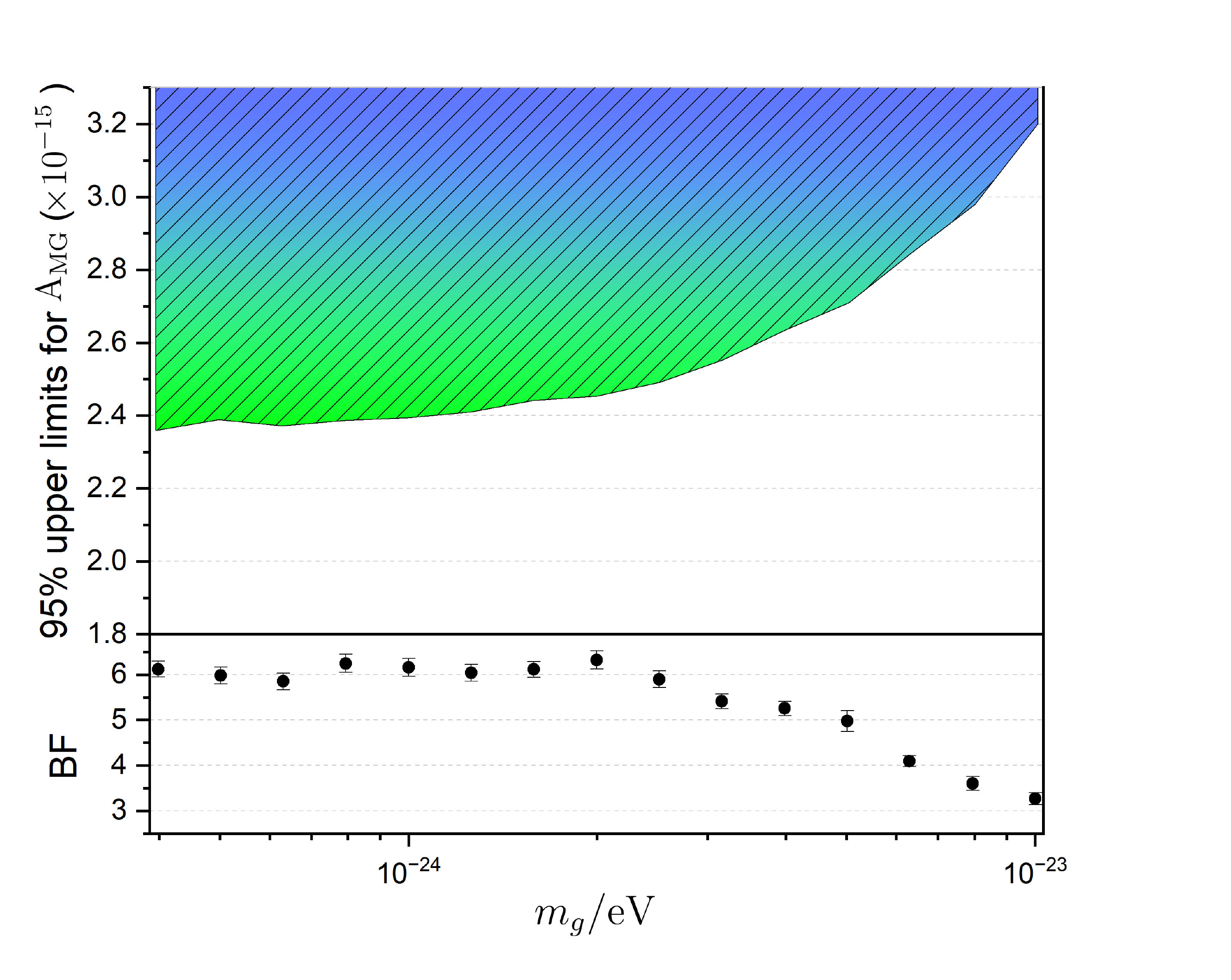}\caption{ \label{A_BF}\textbf{Top panel:} the $95\%$ upper limits on the power spectrum amplitude $A_{\rm{MG}}$ of the SGWB as a function of the graviton mass $m_g$. \textbf{Bottom panel:} the corresponding Bayes factors as a function the graviton mass $m_g$.}
\end{figure}

Within the mass range we probe, we perform a single Bayesian analysis for each given graviton mass shown in \Table{prior} by calculating the Bayes factor between the massive SGWB hypothesis $H_1$ with the correlations of \Eq{orf2} and the UCP hypothesis $H_0$.
The results show that the Bayes factors are larger than 3 but smaller than 7, indicating a positive but no strong evidence \cite{BF} for the SGWB with the dispersion-related correlations. So we put $95\%$ upper limits on the power spectrum amplitude $A_{\rm{MG}}$ for each graviton mass. The upper limits of $A_{\rm{MG}}$ and the Bayes factors for a massive SGWB from the NANOGrav 12.5-year data set are shown in \Fig{A_BF} as a function of the graviton mass. As one can see, the Bayes factors are not large enough to declare the detection for a massive SGWB signal at any certain mass. Meanwhile, the variation trend that the Bayes factor decreases with the increasing mass implies the preference for a lighter or even a massless graviton.

We note that the current GW events from ground-based detectors have put an upper limit of $1.27\times 10^{-23}$~eV on the graviton mass \cite{LIGOScientific:2021sio}. As a comparison, PTAs hold the potential to probe graviton mass lower than $10^{-23}$~eV, hopefully offering constraints complementary to the ground-based detectors.

\textit{Conclusion and Prospects.} PTAs provide a unique probe for gravity theory via the measurement of the spatial correlations of timing residuals induced by an SGWB. In this work, we explore the possibility of an SGWB from massive gravity in the NANOGrav 12.5-year data set by comparing it with the recently reported common-spectrum process. We find no significant evidence for a massive SGWB in the data set, and the Bayes factor prefers a massless graviton. Although we cannot put effective constraints on the graviton mass with the current sensitivity, we place the $95\%$ upper limits on the amplitude of the SGWB for graviton mass smaller than $10^{-23}$~eV as $A_{\rm{MG}}<3.21\times 10^{-15}$. The ruled-out parameter space is shown in \Fig{A_BF}. 

Currently, three major PTA collaborations, i.e., the NANOGrav, the PPTA, and the EPTA, are involved in the effort of SGWB search, and they jointly form the IPTA \cite{Manchester:2013ndt}. Other burgeoning projects, like the Indian PTA (InPTA) \cite{Tarafdar:2022toa}, the Chinese PTA (CPTA) \cite{2016ASPC..502...19L} and the MeerKAT interferometer \cite{Bailes:2020qai}, are joining IPTA collaboration.
With the increasing timespan and the number of pulsars, the sensitivity is expected to improve significantly in the near future. If substantial progress is made in detecting an SGWB with the improved resolution of spatial correlations, we will hopefully place bounds on the graviton mass with PTAs.

\textit{Acknowledgements.}
We thank the referee for very useful comments, and also Xingjiang Zhu and Qiuyue Liang for the helpful conversations.
We acknowledge the use of HPC Cluster of ITP-CAS and HPC Cluster of Tianhe II in National Supercomputing Center in Guangzhou. QGH is supported by the grants from NSFC (Grant No.~12250010, 11975019, 11991052, 12047503), Key Research Program of Frontier Sciences, CAS, Grant No.~ZDBS-LY-7009, CAS Project for Young Scientists in Basic Research YSBR-006, the Key Research Program of the Chinese Academy of Sciences (Grant No.~XDPB15). 
ZCC is supported by the National Natural Science Foundation of China (Grant No.~12247176) and the China Postdoctoral Science Foundation Fellowship No.~2022M710429.

\bibliography{massive_SGWB}
\end{document}